\documentclass[a4paper]{jpconf}

\usepackage{graphicx}
\usepackage{amssymb,amsmath}
\usepackage{color}

\newcommand{\new}[1]{#1}
\begin{document}
\title{Proper motion of the radio pulsar B1727$-$47 and its association with the supernova remnant RCW~114 }

\author{P S Shternin$^1$, M Yu$^2$, A Yu Kirichenko$^{3,1}$, Yu A Shibanov$^{1,4}$, A A Danilenko$^1$, M A Voronkov$^5$ and D A Zyuzin$^1$}

\address{$^1$ Ioffe Institute, St. Petersburg, Russia}
\address{$^2$ National Astronomical Observatory, Chinese Academy of Science, Beijing, China}
\address{$^3$ Universidad Nacional Autonomia de Mexico, Ensenada, Baja California, Mexico}
\address{$^4$ Peter the Great St.~Petersburg Polytechnic University,  St. Petersburg, Russia}
\address{$^5$ CSIRO Astronomy \& Space Science, Epping, New South Wales, Australia}

 \ead{pshternin@gmail.com}

\def\j1731{J1731$-$4744}
\def\b1727{B1727$-$47}

\def\apj{ApJ}
\def\mnras{MNRAS}
\def\nat{Nature}
\def\aap{A\&A}

\begin{abstract}
We report preliminary results of the analysis of the proper motion
of the bright radio pulsar \b1727. Using archival Parkes timing
data, as well as original and archival ATCA interferometry
observations, we, for the first time, constrain the pulsar proper
motion at the level of 148$\pm$11~mas~yr$^{-1}$. The backward
extrapolation of the proper motion vector to the pulsar birth
epoch points at the center of the Galactic supernova remnant
RCW~114 suggesting the genuine association between the two
objects. We discuss the implications of the association and argue
that the distance to the system is less than $1$~kpc. This value
is at least two times lower than the dispersion measure distance
estimates. This suggests that the existing Galaxy electron density
models are incomplete in the direction to the pulsar.

\end{abstract}

\section{Introduction}

Identifying a pulsar (PSR) -- supernova remnant (SNR) association
leads to valuable conclusions on the properties of both
components, such as distances, ages, evolution stages etc.
Although there is a little doubt that each PSR was once born in a supernova
explosion, only a few associations have been identified so far with
high confidence (e.g., \cite{Yao2017ApJ}). Even if the pulsar
position projects on the remnant image, the probability is high
that this is due to a chance coincidence \cite{Gaensler1995MNRAS}.
Therefore each possible association should be considered in detail
separately, and one of the strongest arguments can come from the
measurement of the pulsar proper motion (p.m.)
\cite{Kaspi1998AdSpR}. Here we accomplish this goal for
PSR~\b1727\ and SNR RCW~114.

\section{PSR B1727$-$47 and its proper motion}

PSR B1727$-$47 was among the first pulsars observed since the
beginning of the Pulsar Era. It was discovered in 1968 with the
Molonglo observatory \cite{Large1968Natur} and is one of the
brightest among the young ($<100$~kyr) radio pulsars known. It has
the period $P=0.83$~s,
 the characteristic age
$\tau\equiv P/(2\dot{P})=80$~kyr and the dispersion measure (DM)
of $123$~pc~cm$^{-3}$. According to the widely-used NE2001 model
of the Galactic electron density distribution
\cite{cordes2002astro.ph}, the DM-based distance to the pulsar is
$D=2.7$~kpc. However, the more recent model \cite{Yao2017ApJ}
places PSR~\b1727\ as far as at $5.5$~kpc.

\begin{table}[t]
\caption{PSR B1727$-$47 p.m.\ obtained by various methods
discussed in the text. CP -- by comparing published positions; I
-- interferometry; T -- from interglitch solutions; TN --
\textsf{TEMPONEST} results; F -- full combined solution. Cartesian
components of the p.m.\ vector in RA ($\mu_\alpha$) and Dec.\
($\mu_\delta$) directions and the total p.m.\ value $\mu$ are
given.}\label{tab:proper}
\begin{center}
\lineup
  \begin{tabular}{*{6}l}
  \br
   & \m CP&\m I & \m T &\m TN & \m F \cr
   \mr
   $\mu_\alpha$, mas~yr$^{-1}$ & \m $63\pm4$ &  \m $120\pm 9$ &\m $65\pm13 $ & \m$47\pm14$ & \m$73\pm 5$\cr
   $\mu_\delta$, mas~yr$^{-1}$ & $-92\pm 21$ &  $-162\pm 34$ &$-224\pm28$ &  $-132\pm  37$ & $-127\pm 13$\cr
   $\mu$,\, \  mas~yr$^{-1}$ &\m$111\pm 18$   & \m $202\pm 28$ &\m$232\pm27$ & \m $141\pm  36$
   &\m
$148\pm11$\cr
   \br
  \end{tabular}
  \end{center}
\end{table}

\begin{figure}[ht]
\begin{center}
\begin{minipage}{0.44\textwidth}
\includegraphics[width=\textwidth]{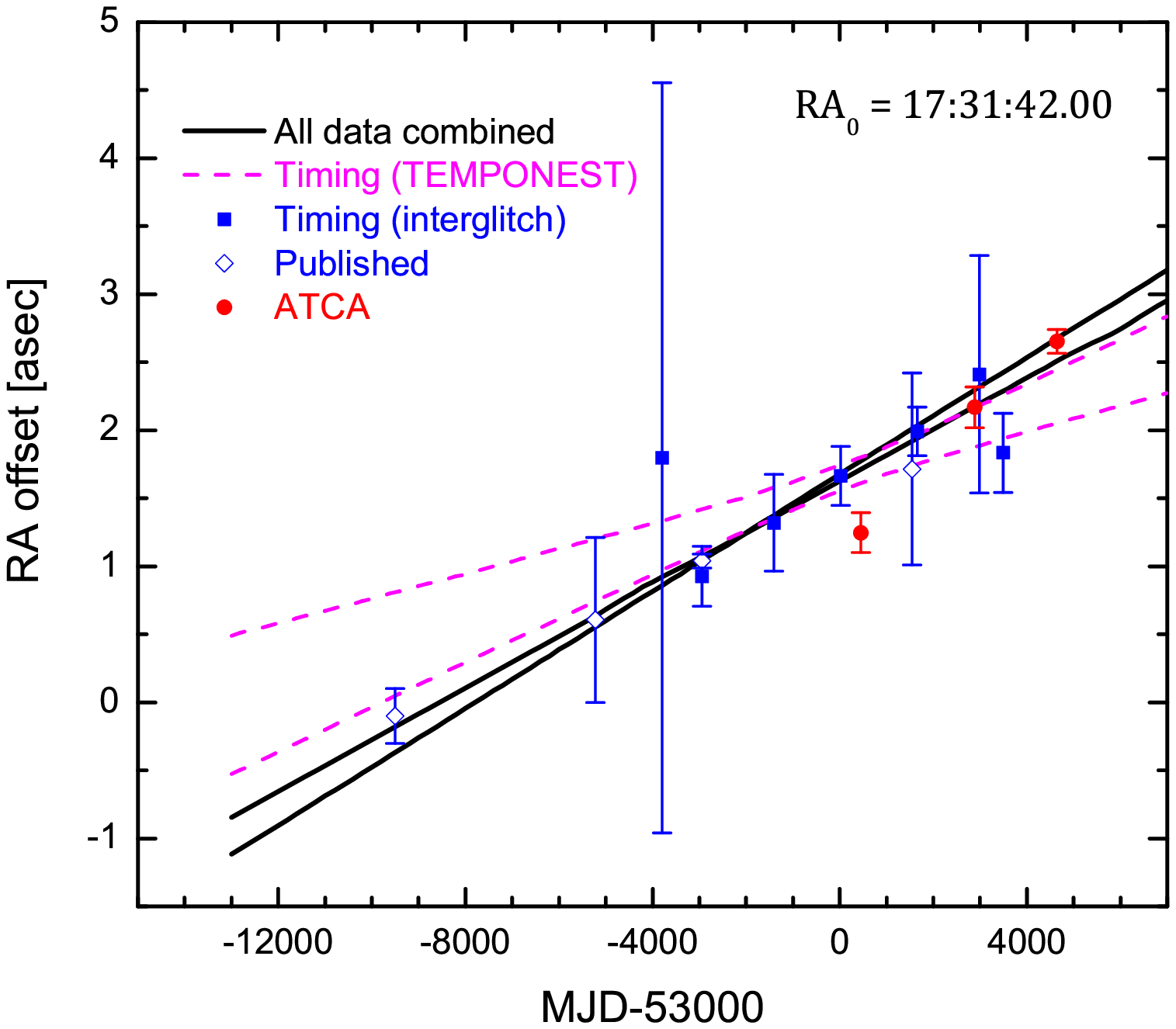}
\end{minipage}\hspace{0.1\textwidth}%
\begin{minipage}{0.44\textwidth}
\includegraphics[width=\textwidth]{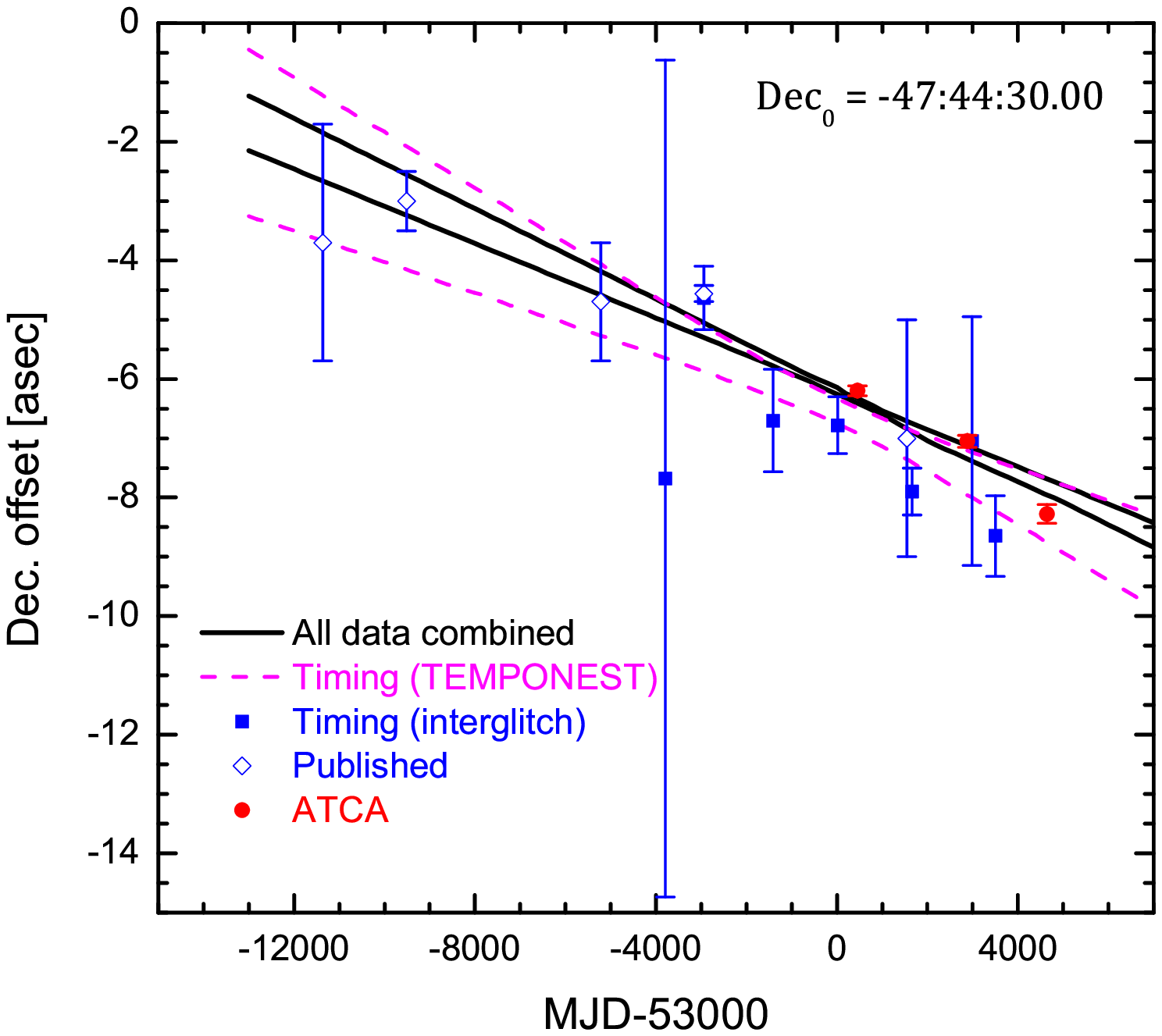}
\end{minipage}
\end{center}
\caption{\label{fig:pos} PSR~\b1727\ positions discussed in this
work. We show relative offsets  in RA ({\it left}) and Dec.\ ({\it
right}) directions from an auxiliary reference point, whose J2000
coordinates are indicated. Open diamonds correspond to previously
published positions \cite{Manchester1983MNRAS,Yu2013MNRAS}. Other
positions are obtained in this work. Blue squares: Parkes timing
solutions for interglitch epoches; red points: ATCA interferometry
positions. Dashed lines bracket the 68 per cent \textsf{TEMPONEST}
credible region and black  lines are the same for all the data are
combined, see text for details.}
\end{figure}

To the best of our knowledge, despite a long observational history, the
p.m.\ of PSR \b1727\ has never been reported.
However, a comparison of the published positions (blue diamonds in
figure~\ref{fig:pos}) suggests that its p.m.\ is significant. The
result of a simple linear fit to these data is given in the first
column in table~\ref{tab:proper}. At $D=2.7$~kpc, the estimated
p.m. value  $\mu^{\rm CP}=111\pm 18$~mas~yr$^{-1}$ leads to a
suspiciously large transverse velocity $v_\perp=1420\pm
230$~km~s$^{-1}$ (note that a typical 2D pulsar velocity is about
250~km~s$^{-1}$ \cite{hobbs2005MNRAS}). Thus the p.m.\ of
PSR~\b1727\ deserves a more detailed study. We have done this
using Parkes timing and ATCA interferometry imaging data.

\subsection{Parkes timing analysis}
We used the processed timing data from the Parkes telescope
archive\footnote{http://data.csiro.au}. We selected the data
obtained from Feb 1993 to Mar 2014. Detailed information on the
observations and filterbank systems is described in
\cite{Yu2013MNRAS}.  For analysis we used 222 times of arrival
(TOAs) spanning 21~yr in total (MJD range 49043--56740). For the
timing analysis we used  the \textsf{TEMPO2}  package
\cite{Hobbs2006MNRASTempo2}. The pulsar is known to be glitching.
Four glitches have been already reported \cite{Yu2013MNRAS}. In more
recent data we identified two new glitches, however the space
limitations preclude us from the discussion of their parameters
here. To measure the p.m., we first constructed timing
solutions for seven inter-glitch periods in our data. The
positions obtained in this way are shown in figure~\ref{fig:pos}
with blue squares and the corresponding p.m. obtained from  these
positions is given in the `T' column in table~\ref{tab:proper}.
The construction of the coherent solution for the full data span
is complicated not only due to the presence of glitches but also
because of the strong timing noise. To get rid of this, we
employed the Bayesian \textsf{TEMPONEST} utility
\cite{Lentati2014MNRAS} which simultaneously optimizes the timing
solution, glitches' parameters and timing noise model. The
obtained p.m.\ is given in the `TN' column of
table~\ref{tab:proper}, and the resulting 68 per cent credibility
pulsar track is shown in figure~\ref{fig:pos} within red dashed
lines.

\subsection{ATCA radio interferometry analysis}
PSR~\b1727\ was observed with ATCA in 2004$-$2005 (project C1323)
and in 2011 November (project C2566). For the C1323 observations,
we used the data set obtained on 2005 March 26, as this session
provides sufficient uv-coverage for accurate position
measurements.
The details of the array configurations and correlator modes can
be found in the ATCA archive. The data were processed and analyzed
in a standard way using the \textsf{MIRIAD} package \cite{miriad}.
The measured pulsar positions revealed a significant shift between
the epochs in a direction compatible with the timing measurements
(figure~\ref{fig:pos}). To verify that the shift is not caused by
some systematic effects, we performed observations
({project CX367) of the pulsar with  ATCA on 2016 September 15
using Director's time. All ATCA positions are shown in
figure~\ref{fig:pos} with red dots, and the corresponding p.m.\
values are given in the `I' column in table~\ref{tab:proper}.

The interferometric and timing observations give qualitatively
similar results. Taking  the weighted mean of all the data
(namely, ATCA positions, older published positions and
\textsf{TEMPONEST} results)   we obtain our final result given in
the last column in table~\ref{tab:proper}. The corresponding
pulsar track uncertainty is constrained in figure~\ref{fig:pos}
within black lines at 68 per cent credibility level.

\begin{figure}[t]
\begin{center}
   \begin{minipage}[h]{0.35\textwidth}
     \caption{H$\alpha$ image image of the RCW~114 field. The apparent radius of RCW~114 is about 2$^{\circ}$.
     PSR~\b1727\ position at the present epoch is marked by the circle, and the 68 per cent credibility region for the pulsar track extrapolated backwards is
  shown by dashed lines.
 }\label{fig:RCW114}
 \end{minipage}
 \hspace{0.05\textwidth}%
 \begin{minipage}{0.35\textwidth}
    \includegraphics[width=\textwidth]{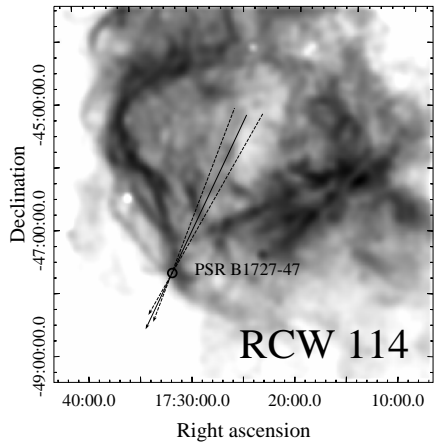}
    \end{minipage}
\end{center}
\end{figure}

\section{Supernova remnant RCW~114}

The backward extrapolation of the measured p.m.\ vector
to the PSR~\b1727\ birth epoch (using $\tau=80$~kyr\new{; the trajectory distortion due to effect of Galactic potential is negligible for this age}) points
roughly at the center of the nearby H$\alpha$ nebula RCW~114, also
known as SNR~G343.0--6.0 \cite{Green2014}
(figure~\ref{fig:RCW114}). This strongly suggests a genuine
association between the two objects and independently constrains
the system age at $\sim 60-80$~kyr.
The SNR nature of RCW~114 is supported by a recent discovery of
C[IV] emission from the nebula \cite{Kim2010ApJ}. This detection
also allowed to place an upper limit on the distance to SNR,
$D_{\rm SNR}\lesssim 1$~kpc \cite{Kim2010ApJ}. On the other hand,
detailed spectroscopic analysis of the stars projected on RCW~114
suggests $D_{\rm SNR}>0.5$~kpc \cite{Welsh2003}. The distance
range of $0.5-1$~kpc is plausible if the SNR is at the snowplough
stage \cite{Cioffi1988ApJ}. Using expressions from
\cite{Cioffi1988ApJ} for typical explosion energy of
$10^{51}$~erg,  ambient density of $\sim 1$~cm$^{-3}$ and $D_{\rm
SNR}=0.7$~kpc, we estimated the expansion velocity of
$96$~km~s$^{-1}$, which is in accord with the velocity of the
faint outer filament of $-80$~km~s$^{-1}$ reported in
\cite{Meaburn1991}.

\section{Conclusions}

\begin{itemize}
  \item Based on radio timing and interferometric observations, we
  find the  proper motion of PSR~\b1727\ of 148$\pm$11~mas~yr$^{-1}$. For the DM-based
  distance of $2.7$~kpc ($5.5$~kpc) it leads to extremely
  high  $v_\perp\approx 1900$~km~s$^{-1}$ (3900~km~s$^{-1}$).
  \item The proper motion vector points at the center of the
  SNR RCW~114, suggesting SNR--PSR association. In this case the
   most plausible distance to the system is $\sim 0.7$~kpc and the
   p.m.-based age is $60-80$~kyr, compatible
   with  the pulsar characteristic age. \new{The relatively low distance value can be checked independently by the parallax measurements with very long baseline interferometry.}
   \item Assuming $D=0.7$~kpc, we find
   $v_\perp\approx 500$~km~s$^{-1}$, which is well in line with the 2D pulsar velocity
   distribution \cite{hobbs2005MNRAS}.
   This result suggests that the    existing models of the electron density in the Galaxy require some tuning in the
   direction to PSR~\b1727. 
   \new{For instance, inclusion of a nearby `clump' of enhanced electron density \cite{cordes2002astro.ph} in the model  can reconcile the high DM value and the low distance inferred here. }
\end{itemize}
\vspace{-0.3cm}

\ack
We thank G~B Hobbs, L Lentati and R~N Manchester for
discussions and the referee S Zane and the editor G~G Pavlov for useful comments. The work was supported by the Russian Science
Foundation, grant 14-12-00316. The Australia Telescope Compact
Array is part of the Australia Telescope National Facility which
is funded by the Australian Government for operation as a National
Facility managed by CSIRO. This paper includes archived data
obtained through the Australia Telescope Online Archive
(http://atoa.atnf.csiro.au).

\vspace{-0.2cm}
\section*{References}

\providecommand{\newblock}{}


\end{document}